\documentstyle[mnextra]{mn} 
\input{epsf}
\def \etal {et al.\ }

\nobrackets     

\newcommand{\galf}{\textsc{galform}}

\begin{document}

\title[N-body \galf]{Galaxy formation using halo merger histories taken from N-body simulations}

\author[J. C. Helly et al.] {
\parbox[h]{\textwidth}{John C. Helly$^1$, Shaun Cole$^1$, Carlos S. Frenk$^1$,
Carlton M. Baugh$^1$, Andrew Benson$^2$, Cedric Lacey$^1$}
\vspace*{6pt} \\ 
$^1$Department of Physics, University of Durham, Science Laboratories, South 
Road, Durham DH1 3LE, United Kingdom \\
$^2$California Institute of Technology, MC105-24, Pasadena CA 91125, USA \\
}
\maketitle
 
\begin{abstract}
We develop a hybrid galaxy formation model which uses outputs from an N-body simulation to follow the merger histories (or ``merger trees'') of dark matter halos and treats baryonic processes, such as the cooling of gas within halos and subsequent star formation, using the semi-analytic model of Cole \etal We compare this hybrid model to an otherwise identical model which utilises merger tree realisations generated by a Monte-Carlo algorithm and find that, apart from the limited mass resolution imposed by the N-body particle mass, the only significant differences between the models are due to the known discrepancy between the distribution of halo progenitor masses predicted by extended Press Schechter theory and that found in N-body simulations. We investigate the effect of limited mass resolution on the hybrid model by comparing to a purely semi-analytic model with greatly improved mass resolution. We find that the mass resolution of the simulation we use, which has a particle mass of $1.4 \times 10^{10} \it{h}^{-1} \rm{M}_{\odot}$, is insufficient to produce a reasonable luminosity function for galaxies with magnitudes in the $\rm{b_J}$ band fainter than -17.
\end{abstract}

\section{Introduction}
\label{sec:intro}

Hierarchical models of galaxy formation must describe both the growth and collapse of density perturbations to form dark matter halos and the baryonic processes which lead to the formation of stars. Despite uncertainty as to the exact nature of the dark matter itself, the formation and evolution of dark matter halos appears to be reasonably well understood. The two main approaches to this problem are direct numerical simulations and analytic techniques such as the Press-Schechter theory (\cite{ps74}). Encouragingly, the mass functions of dark matter halos predicted using these very different approaches are found to agree to within 50\% (\cite{gross98,governato99,j2001}). The analytic model described by Sheth \& Tormen (\shortcite{st01}) based on the assumption that objects collapse ellipsoidally rather than spherically achieves even better agreement with N-body simulations. Mo \& White (\shortcite{mw02}) present halo abundances from this and several other models.

This understanding of the hierarchical build up of structure provides the starting point for semi-analytic models of galaxy formation, which  attempt to follow the development of galaxies from primordial density fluctuations. In semi-analytic models, merger histories for dark matter halos may be taken directly from dark matter simulations (e.g. \cite{kauffmann99,vk99}).  Alternatively, extensions to the Press-Schechter theory which predict the conditional halo mass function (\cite{bond91,bower91}) and halo survival times, formation times and merger rates (\cite{lc93}) may be used to construct realisations of merger histories for individual halos. Simple analytic modelling is then used to follow the evolution of the baryonic component, including prescriptions for processes such as star formation and its possible effects on the remaining gas. Semi-analytic models (e.g. \cite{cole91,ls91,wf91,cole94,sp99,cole2k}) have successfully reproduced many observable properties of galaxies, such as the local field galaxy luminosity function and distributions of colour and morphology. When combined with N-body simulations, semi-analytic models have also successfully reproduced galaxy clustering properties (e.g. \cite{governato98,kauffmann99,b2000,wechsler2001}).

Semi-analytic models utilising merger trees generated using algorithms based on the extended Press-Schechter (EPS) formalism have two closely related advantages over models which take merger histories from N-body simulations. Creating Monte-Carlo realisations of merger trees for a set of halos generally requires fewer computing resorces than carrying out an N-body simulation of a similar number of halos. In both cases, improving the mass resolution increases the computational load, but since the load is much less in the Monte-Carlo case, significantly better mass resolution may be achieved. Methods based on the Press-Schechter theory, however are only applicable to initially Gaussian fluctation fields. N-body simulations, on the other hand, have the advantage that the non-linear evolution of density fluctuations is followed in complete generality, without the need for any of the assumptions involved in creating EPS merger trees. 

There are advantages to both of these methods, and which is more appropriate depends on the problem being addressed. In this paper we investigate the effects of the choice of merger trees on the predictions of one particular semi-analytic model. We describe a new method of extracting merger trees from an N-body simulation and incorporate these merger trees into a semi-analytic galaxy formation model based on that of Cole \etal (\shortcite{cole2k}). We compare the predictions of this model to those of a similar model utilising Monte-Carlo realisations of halo merging histories. In order to identify the reasons for the discrepancies that we find, we determine the changes that must be made to the Monte-Carlo model to reproduce the N-body results.

The use of N-body merger trees in semi-analytic models allows a halo-by-halo comparison between the semi-analytic treatment of baryonic processes, such as gas cooling, and direct numerical simulations of galaxy formation. In a companion paper (\cite{paperII}) we carry out such a comparison between a ``stripped down'' version of the semi-analytic model described in this paper and a smoothed particle hydrodynamics simulation of a cosmological volume.

This paper is laid out as follows. In Section \ref{sec:nbodytrees} we explain how we obtain merger trees from an N-body simulation. In Section \ref{sec:nbody-galform} we investigate the effect on our semi-analytic model of utilising merger trees  derived from N-body simulations rather than Monte-Carlo realisations. In Section \ref{sec:conclusions} we present our conclusions.

\section{Extracting Merger Trees}
\label{sec:nbodytrees} 

We now present the method we used to calculate the merger histories of dark matter halos identified in an N-body simulation. The simulation, which will be referred to as the GIF simulation, was carried out by the Virgo Consortium using a parallel adaptive particle-particle/particle-mesh ($\rm{AP^3M}$) code known as Hydra (\cite{ctp95,pc97}) as part of the GIF project. The simulation assumes the $\rm{\Lambda CDM}$ cosmology with mean mass density parameter $\Omega_0=0.3$, cosmological constant $\Lambda_0=0.7$ in units of $3H_0^2/c^2$, power spectrum shape parameter $\Gamma=0.21$, present day rms linear fluctuation amplitude in $8h^{-1}\rm{Mpc}$ spheres $\sigma_8=0.90$, and Hubble constant $h=0.7$ in units of $100\rm{km s^{-1} Mpc^{-1}}$. It contains $256^3$ dark matter particles each of mass $1.4 \times 10^{10} {\it{h}}^{-1}\rm{M}_{\odot}$ in a box of side $141.3 h^{-1}\rm{Mpc}$. The gravitational softening length in the simulation is $30h^{-1}\rm{kpc}$ at $z=0$. This simulation is described in more detail by Jenkins \etal (\shortcite{j98}), where it is referred to as $\Lambda\rm{CDM2}$,  and by Kauffmann \etal (\shortcite{kauffmann99}). While halo catalogues and merger trees based on this simulation are publically available, here we make use of only the simulation outputs themselves and construct merger trees using a somewhat different algorithm to that of Kauffmann \etal We use 44 output times from the simulation which are spaced equally in $\log_{10} (1+z)$ between $z=0$ and $z \sim 20$.

\subsection{Identifying Halos}
\label{sec:groups} 

In order to construct merger histories for dark matter halos in an N-body simulation, a catalogue of halos must be produced for each simulation output using a group finding algorithm. The algorithm used here is the ``friends of friends'' (FOF) method of Davis \etal (\shortcite{davis85}), which simply links together any particles with separations less than the linking length $b$, usually expressed in terms of the mean interparticle separation. Given sufficiently large numbers of particles in each object, the FOF algorithm finds regions bounded by a surface of constant density. The density threshold is proportional to $1/b^3$.

The FOF approach has the advantage that it imposes no constraints on the geometry of the halos identified, but it may occasionally artificially join two nearby halos if a transient ``bridge'' of a few particles forms between them. It will be seen in Section~\ref{sec:trees} that this can cause problems when attempting to generate merger trees using FOF group catalogues, and a method of identifying and splitting artificially joined halos is described in Section~\ref{sec:trees}.

The usual choice for the linking length in cosmologies with $\Omega=1$ is $b=0.2$ (e.g. \cite{lc94}), which identifies halos with a mean density similar to that predicted by the top hat spherical collapse model (\cite{cl96}). However, in cosmologies with $\Omega<1$ there is no rigorous justification for any particular choice. Here, we choose to set $b=0.2$ at all redshifts as in the $\Omega=1$ case. See Eke, Cole \& Frenk (\shortcite{ecf96}) and Jenkins \etal (\shortcite{j2001}) for further discussion.

The other parameter needed by the FOF algorithm is the minimum number of particles, $N_{\rm{min}}$, required to constitute a group. It is important that $N_{\rm{min}}$ be as small as possible, since detailed merger trees can only be obtained for halos much larger than the smallest resolvable group. Kauffmann \etal (\shortcite{kauffmann99}) found that in their simulations groups as small as 10 particles are dynamically stable systems and that for 95\% of these groups, 80\% of the particles remain in the same group at subsequent times.

We therefore identify halos using a linking length $b=0.2$ at all redshifts, with a minimum group size of ten particles. The resulting catalogues may still contain some groups which consist of unbound particles which happen to be close together at this particular timestep. To remove these, we follow Benson \etal (\shortcite{b2001b}) and calculate the total energy of each group. Unbound groups are not immediately discarded, because they may only be unbound due to the presence of a small number of fast moving particles. The binding energy of each particle is calculated, and the least bound particle is removed from the group. This is repeated until the group becomes bound. If half of the particles are removed or the group is reduced to less than $N_{\rm{min}}$ particles we discard it. Up to 5\% of all groups are discarded, with a similar number of groups  being reduced in mass by this procedure. The affected groups generally consist of around 10-20 particles.

We use the procedure described above to generate halo catalogues for 44 simulation outputs between redshifts $z=20$ and $z=0$, spaced approximately evenly in $\log_{10} (1+z)$.

\subsection{Constructing N-body Merger Trees}
\label{sec:trees} 

In an idealised picture of the process of hierarchical structure formation (e.g. Press-Schechter theory), dark matter halos may increase in mass by mergers, but cannot lose mass. Consequently, any halo identified in a simulation prior to the final output time should still exist at subsequent output times, although it may have become subsumed within a larger halo through a merger. In any case, the constituent particles of the original halo should still all be members of a single group. It should therefore be possible to identify each halo in the simulation as a progenitor of a single halo at the next output time. 

In practice there are several ways in which a halo can lose particles. Halos may be disrupted by tidal forces caused by other nearby halos. The masses of simulated halos can also fluctuate because the FOF algorithm imposes a somewhat arbitrary boundary on the halo and outlying particles which are considered group members at one timestep may lie just beyond the boundary at the next timestep. 

\begin{figure}
\epsfxsize=8.5 truecm \epsfbox{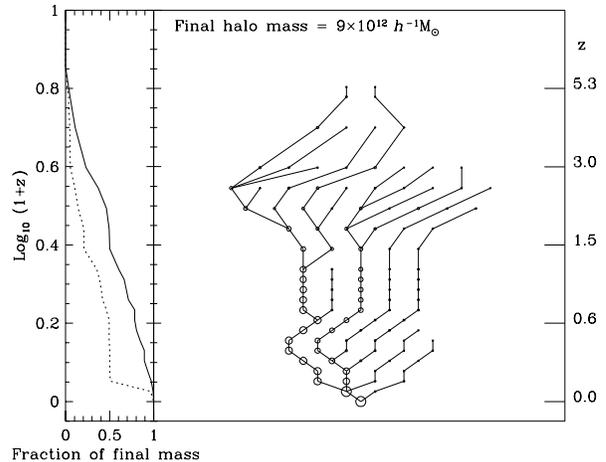}
\caption{An example of a merger tree obtained from the GIF simulation for a halo of mass $9 \times 10^{12}h^{-1}\rm{M_{\odot}}$ at redshift $z=0$. Each circle represents a dark matter halo identified in the simulation, the area of the circle being proportional to the halo mass. The vertical position of each halo on the plot is determined by $\log_{10} (1+z)$ at the redshift at which it exists, the horizontal positioning is arbitrary. The solid lines connect halos to their progenitors. The solid line in the panel on the left-hand side shows the fraction of the final mass contained in resolved progenitors as a function of redshift. The dotted line shows the fraction of the final mass contained in the largest progenitor as a function of redshift.}
\label{fig:tree}
\end{figure}

The technique we use to determine merger histories is intended to take into account this uncertainty in the definition of a halo and a possible loss of particles. First, we consider two adjacent output times from the simulation, $t_1$ and $t_2$, where $t_1<t_2$. Each halo at time $t_1$ is labelled as a progenitor of whichever halo at time $t_2$ contains the largest fraction of its particles. This process is repeated for all pairs of adjacent output times. It is then straightforward to trace the merger history of each halo which exists at the final output time. Fig.~\ref{fig:tree} shows an example of a merger tree created in this way for a halo with a final mass of about $9 \times 10^{12} \it{h}^{-1}\rm{M}_{\odot}$, or around 700 particles.

In the semi-analytic model used here, galaxies are assumed to form at the centres of dark matter halos, so the centre of each halo in the merger tree must be defined. We choose to follow Kauffmann \etal (\shortcite{kauffmann99}), who identified the most bound dark matter particle as the position of any galaxy which forms in the halo. We define the binding energy of a particle as the sum of its kinetic energy and the gravitational potential energy due to the other particles in the halo. This approach differs from that of Benson \etal (\shortcite{b2001a}), who associated the central galaxy in a halo with the centre of mass. Once a galaxy forms it is assumed to follow this particle until the parent halo merges with another halo and dynamical friction, calculated as described in Cole \etal (\shortcite{cole2k}), causes the galaxy to merge with the central galaxy of the new halo. We therefore check that the most bound particle of a halo remains a member of the same halo as the majority of the halo's constituent particles at the next output time. If this is not so, we choose the most bound particle from those which \emph{are} in the correct halo at the later output time. This problem generally only occurs in smaller halos which may be easily disrupted.

During the construction of the merger trees, we also attempt to deal with the problem mentioned in Section \ref{sec:groups} --- the possibility that nearby halos may be artificially linked by the FOF algorithm. The problem occurs if two halos become temporarily linked by a transient ``bridge'' of particles which causes the FOF group finder to consider them as a single, large group. When the bridge is later broken, the group splits, leaving the two original halos. Our tree building method would identify the large, joined group as a progenitor of the larger of the two final groups.

These situations are identified by looking for groups at the earlier time $t_1$ whose particles are shared between two or more groups at the subsequent output time $t_2$. This indicates that between times $t_1$ and $t_2$ the group has split into smaller groups which we refer to here as ``fragments''.
 
We split such spuriously joined groups into one new group for each fragment which contains more than $N_{\rm{min}}$ of its constituent particles. Particles belonging to one of these fragments at time $t_2$ are assigned to the corresponding new group at the earlier time $t_1$. Particles belonging to no fragment, or to a fragment with fewer than $N_{\rm{min}}$ particles from the joined group, are assigned to the new group corresponding to the fragment ``closest'' to their position at time $t_1$. The separations used are weighted by a factor $M^{-1/3}$ to account for the spatial extent of the groups, where $M$ is the mass of the fragment.

The splitting procedure is first carried out for halos at the penultimate timestep and then repeated for each earlier output time in order of increasing redshift. For each timestep a modified group catalogue is produced, which is then used to determine whether any halos at the previous timestep need to be split. This ensures that if any bridge between a pair of halos persists for more than one timestep the halos are split at each timestep where the bridge exists. 

\subsection{Mass Conservation}
\label{sec:mass} 

In the \galf\ semi-analytic model of Cole \etal (\shortcite{cole2k}), halos may gain mass through mergers with other halos. The mass of a halo always increases with time, and the difference between the mass of a halo and the sum of the masses of its progenitors is due to the accretion of small, unresolved dark matter halos.

The N-body merger trees may contain halos which decrease in mass from one timestep to the next for the reasons described in Section \ref{sec:trees} --- the nature of the definition of a halo imposed by the FOF group finder and the possibility of disruption by tidal forces. Consequently, a halo in a N-body merger tree may be somewhat \emph{less} massive than its progenitors. In the \galf\ model this corresponds to the unphysical situation where a negative amount of mass is accreted in the form of sub-resolution halos.

\begin{figure*}
\epsfxsize=18.0 truecm \epsfbox{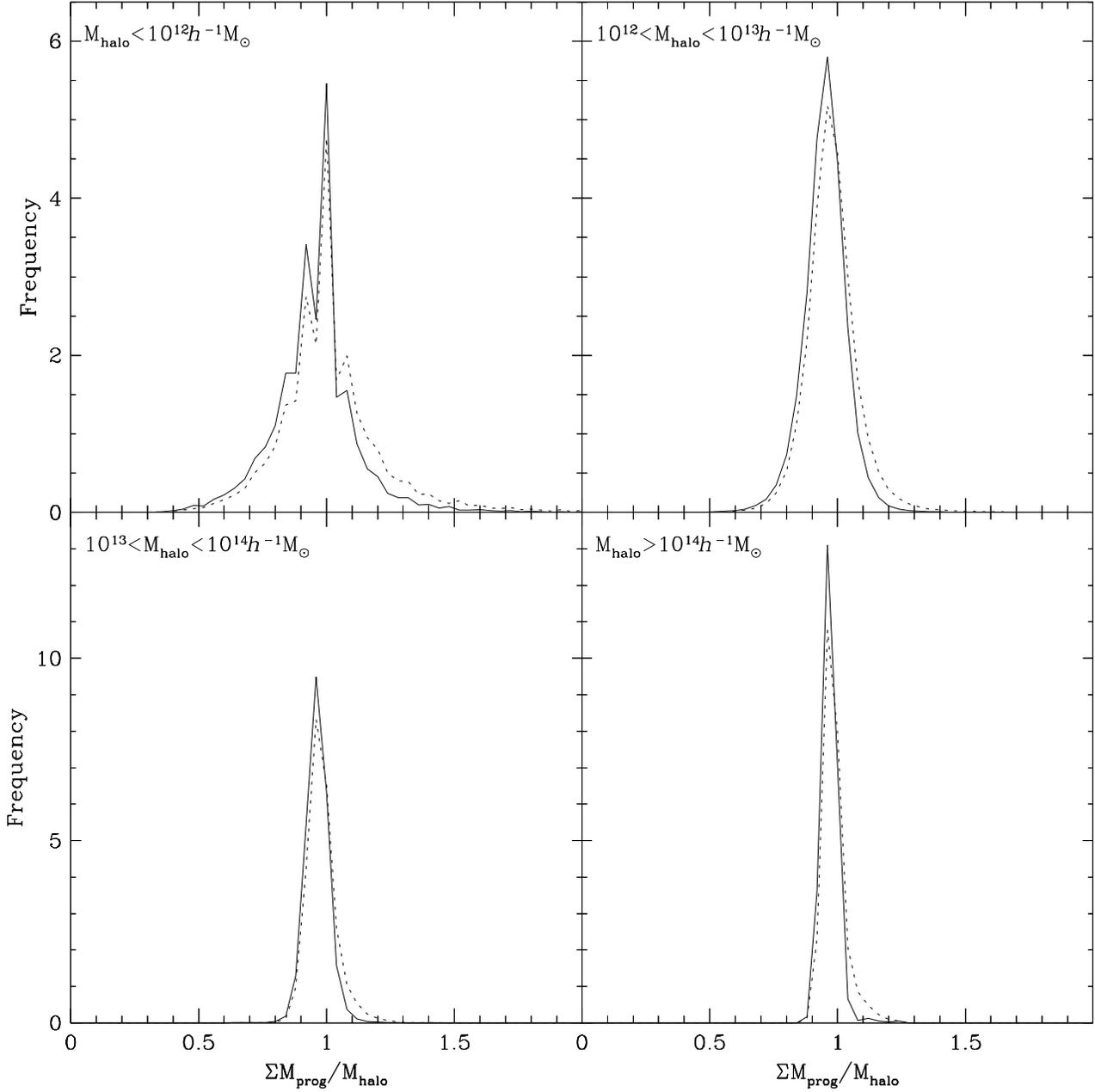}
\caption{The solid lines show the distribution of the ratio of the total mass of the immediate progenitors of a halo, $\rm{\Sigma M}_{\rm{prog}}$, to the mass of the halo at the next timestep, $\rm{M}_{\rm{halo}}$. Each panel shows the distribution of $\rm{\Sigma M}_{\rm{prog}}/\rm{M}_{\rm{halo}}$ for halos in the mass range shown at the top of the panel. The dotted lines show the distribution of $\rm{\Sigma M}_{\rm{prog}}/\rm{M}_{\rm{halo}}$ if $\rm{\Sigma M}_{\rm{prog}}$ is evaluated after the progenitors have been increased in mass to at least the total mass of \emph{their} progenitors. Where this ratio is greater than 1, it is the factor by which $\rm{M_{halo}}$ must be changed to ensure mass conservation if we choose to add mass.}
\label{fig:masscons}
\end{figure*}

The solid lines in Fig.~\ref{fig:masscons} show the distribution of the ratio $\rm{\Sigma M_{prog}/M_{halo}}$, where $\rm{M_{halo}}$ is the mass of a halo and $\rm{\Sigma M_{prog}}$ is the total mass of the immediate progenitors of the halo, which exist at the previous timestep. Halos at all timesteps (other than the first) are included. If these merger trees had been created using the technique of Cole \etal (\shortcite{cole2k}), then this ratio would always be less than one. It can be seen from Fig.~\ref{fig:masscons} that for halos less massive than about $10^{12}h^{-1}\rm{M_{\odot}}$ the total mass in the progenitors can occasionally exceed the mass of the halo they form at the next timestep by up to 50\%. More massive halos are less affected, but there are still rare instances where the largest halos have progenitors with masses 5-10\% greater than the mass of the halo. 

Mass conservation can be forced on the N-body merger trees by simply adjusting the masses of some of the halos. Two opposite approaches to the problem are possible. Mass can be added to those halos which are less massive than their progenitors, or mass can be removed from the progenitors themselves. In order to show that the changes made to the halo masses have little effect on the semi-analytic model, we create merger trees using both methods.

Enforcing the conservation of mass in merger trees by adding mass is relatively straightforward. If a halo is less massive than its progenitors, its mass is increased to match that of the progenitors. The halo may, in turn, be a progenitor of a later halo which may now become less massive than its own progenitors. This later halo's mass will then also be increased. Changes made to halo masses at early times may therefore propagate to later times. 

Similarly, if mass is removed from a halo to force conservation of mass, it may become less massive than its progenitors and reductions in mass could then propagate to earlier times. We attempt to remove mass in such a way as to minimize the effects on earlier halos. Each halo has a certain amount of ``excess'' mass beyond that of its progenitors, which was accreted over the last timestep in the form of sub-resolution objects. This mass, if it exists, may be removed without the change propagating to earlier halos. When a halo which is less massive than its progenitors is found, mass is first removed from the excess mass of the largest progenitor. If still more mass must be removed, it is taken from the excess mass of the other progenitors in decreasing order of mass. If all of the excess mass of the progenitors is removed and yet more mass needs to be taken away, the masses of all of the progenitor halos are simply scaled down by a constant factor.

The dotted lines in Fig.~\ref{fig:masscons} show the sizes of the changes we are forced to make when we enforce mass conservation by adding mass to halos. These lines show the distribution of the ratio $\rm{\Sigma M}_{\rm{prog}}/\rm{M}_{\rm{halo}}$ if $\rm{\Sigma M}_{\rm{prog}}$ is evaluated after the progenitors of the halo at all previous timesteps have been made at least as massive as their own progenitors. $\rm{M}_{\rm{halo}}$ is still the original halo mass. Where this ratio exceeds 1, it is the factor by which $\rm{M_{halo}}$ must be scaled to ensure that the halo is at least as massive as its progenitors. It can be seen that the required changes to individual halos are generally small, and adjustments are required much less frequently in well resolved halos. However, the masses of a minority of halos are affected quite significantly and it is necessary to show that these changes do not affect the galaxy population predicted by the semi-analytic model. The algorithms described above are two opposite ways of dealing with the problem of mass conservation in the merger trees. While artificially altering the halo masses is clearly not ideal, if, as is the case, both methods produce very similar results when the merger trees are fed into the semi-analytic model we can then conclude that the changes we have made are insignificant. This comparison is carried out in Section \ref{sec:effect_mass}.
 
\section{Comparison between GALFORM and N-body GALFORM}
\label{sec:nbody-galform}

In this section we describe our semi-analytic model, indicating how it differs from the model of Cole \etal (\shortcite{cole2k}) on which it is based. We also explain how merger trees obtained from a simulation may be incorporated into the model.

\subsection{The N-body GALFORM model}
\label{sec:galform}

We use the \galf\ semi-analytic model to treat the process of galaxy formation within the dark matter halos in the GIF simulation. The model is described in detail by Cole \etal (\shortcite{cole2k}) so here we present only a brief description of features that are important to this work. The original model of Cole \etal will be referred to as ``standard \galf'', and the version using merger trees taken from a simulation will be referred to as ``N-body \galf''.

The starting point for the standard \galf\ model is a set of merger trees created using a Monte-Carlo technique. The history of each halo is divided into a number of discrete timesteps. Extended Press-Schechter theory is used to estimate the probability that a halo ``fragments'' into two progenitors when a step back in time of size $\delta t$ is taken. The masses of the fragments are chosen at random from a distribution consistent with extended Press-Schechter theory. Halos are repeatedly split in this way to create merger trees. A mass resolution limit is imposed on the merger trees, below which progenitors are considered to be material acquired through continuous accretion. The mass resolution is normally set sufficiently high that the results of interest are not sensitive to its value. In the N-body \galf\ model, we replace these merger trees with those calculated directly from the GIF simulation as described in Section \ref{sec:trees}. The mass resolution limit is then determined by the mass of the smallest halo which can be resolved in the simulation.

The dark matter halos in the merger tree are assumed to be spherically symmetric with the radial density profile of Navarro, Frenk \& White (\shortcite{nfw96}, \shortcite{nfw97}):\begin{equation} \rho (r) \propto \frac{1}{ r /r_{\rm{NFW}} ( r/r_{\rm{NFW}} + 1)^2 }, \end{equation} where $r_{\rm{NFW}}$ is the scale radius of the halo and is related to the concentration parameter, $c$, defined by Navarro, Frenk \& White (\shortcite{nfw97}) through $r_{\rm{NFW}}=r_{\rm{virial}}/c$, where $r_{\rm{virial}}$ is the virial radius of the halo. The concentration parameter is set using the method described in the appendix of the same paper. We do not allow for any scatter in the concentration parameter as a function of halo mass.

Our treatment of the cooling of gas within halos is identical to that of Cole \etal (\shortcite{cole2k}). Initially, the amount of gas in each halo is taken to be equal to the mass of the halo times the universal baryon fraction. The gas is assumed to be shock-heated to the virial temperature of the halo when it forms. We assume that the radial density profile of the gas is given by \begin{equation} \rho _{\rm{gas}} (r) \propto 1 / ( r^2 + r ^2 _{\rm{core}} ), \label{eqn:rhogas} \end{equation} where the core radius is given by $r_{\rm{core}} / r_{\rm{NFW}} \approx 1/3 $ in accordance with the simulations of Navarro \etal (\shortcite{nfw95}). This core radius is allowed to grow with time from an initial value, $r_{\rm{core}}^0$, as gas is removed by cooling in order to maintain the same gas density at the virial radius. This ensures that the pressure at the virial radius, which would be maintained by shocks from infalling material, remains unchanged. 

To determine the rate at which gas can cool and form a disk at the centre of the halo, the cooling time of the gas is calculated as a function of radius using the cooling function of Sutherland \& Dopita (\shortcite{sd93}). Gas which has had time to cool and fall to the centre of the halo is added to the disk where it is available to form stars.

When halos merge, the most massive galaxy becomes the central galaxy in the new halo. The resolution of the simulations used here is insufficient to follow the evolution of substructure within the dark matter halos. Instead, the dynamical friction time scale, as defined by Lacey \& Cole (\shortcite{lc93}), is used to determine when each satellite will merge on to the central galaxy. It should be noted at this point that the orbital parameters used to determine the dynamical friction time for each galaxy are assigned at random from a distribution consistent with the numerical results of Tormen (\shortcite{tormen97}), even when using merger trees obtained from the simulation. 

\subsection{Parameters in the N-body GALFORM model}
\label{sec:galparams}

The \galf\ semi-analytic model requires a number of parameters to be specified, which can be divided into three categories. There are numerical parameters, parameters describing the background cosmology and parameters which describe the physical model of galaxy formation.

The numerical parameters are the mass resolution, $M_{\rm{res}}$, the number of timesteps in the merger tree and the starting redshift. In the N-body \galf\ model these are all constrained by the properties of the simulation used to obtain the merger trees. The mass resolution is the mass of the smallest halo which our group finding algorithm can resolve, there is one timestep for each simulation output and the starting redshift is the redshift of the first output. The cosmological parameters $\Omega_0$, $\Lambda_0$, $h$, $\sigma_8$, $\Gamma$ and, in the case of a simulation with a baryonic component, $\Omega_b$, are also fixed by the simulation. 

The remaining parameters allow us to vary the treatment of the processes involved in galaxy formation. The parameters we are interested in are:
\begin{itemize}
\item{$r_{\rm{core}}^0$: the initial size of the core in the radial gas density profile, specified in terms of $r_{\rm{NFW}}$ (see eqn.~\ref{eqn:rhogas}).}
\item{The evolution of $r_{\rm{core}}$ with time. The radius $r_{\rm{core}}$ may be a fixed fraction of $r_{\rm{NFW}}$ or it may be allowed to increase with time as described in Section~\ref{sec:galform}}
\item{$f_{\rm{df}}$: A factor by which the dynamical friction time scale for a satellite galaxy, which is used to determine when the galaxy merges with the central galaxy of the halo, may be scaled. Increasing $f_{\rm{df}}$ reduces the rate at which galaxy mergers occur within halos.}
\end{itemize}

The other parameters in the model are the same as those in the reference model of Cole \etal (\shortcite{cole2k}), with the following minor changes: $v_{\rm{hot}}=250\rm{km s^{-1}}$ and $f_{\rm{ellip}}=0.5$. The parameter $v_{\rm{hot}}$ determines the efficiency with which energy injection from supernovae and young stars reheats and ejects cold gas from galactic disks. The parameter $f_{\rm{ellip}}$ is used to decide the outcome of mergers between central and satellite galaxies. If the ratio of the mass of the satellite to the mass of the central galaxy is greater than $f_{\rm{ellip}}$, any gas in the disks of the two galaxies is converted into stars and an elliptical galaxy is produced. If the ratio is smaller than $f_{\rm{ellip}}$, any stars present in the satellite are added to the bulge of the central galaxy and any gas is added to the disk. These changes to the Cole \etal model are required to obtain a realistic luminosity function at $z=0$ with the higher baryon density, $\Omega_{\rm{b}}=0.038$, which we use here.

Our prescription for star formation differs slightly from that of Cole \etal In our model, the time scale for star formation is given by \begin{equation}\tau_*=\tau^0_* (V_{\rm{disk}}/200\rm{km\,s^{-1}})^{\alpha_*}, \label{eqn:taustar}\end{equation} where $V_{\rm{disk}}$ is the circular velocity of the galaxy disk and the time scale, $\tau^0_*$, is set to 3Gyr. We set $\alpha_{*}=-2.5$. The way $\tau_{*}$ scales with redshift in this model results in reduced star formation and more gas rich mergers at high redshift and has been shown (\cite{lacey02}) to better reproduce the properties of SCUBA and Lyman break galaxies. Kauffmann \& Haehnelt (\shortcite{kh2000}) also find that a star formation scheme with an increased star formation timescale at high redshift is required to reproduce observations of damped Ly$\alpha$ absorption systems and the increase in number density of bright quasars from $z=0$ to $z=2$. It should also be noted that, for the purposes of this comparison, the details of our star formation prescription are not critical, since the same scheme is used in both the standard and N-body \galf\ models.

\subsection{Effects of mass conservation}
\label{sec:effect_mass}

\begin{figure*}
\epsfxsize=18.0 truecm \epsfbox{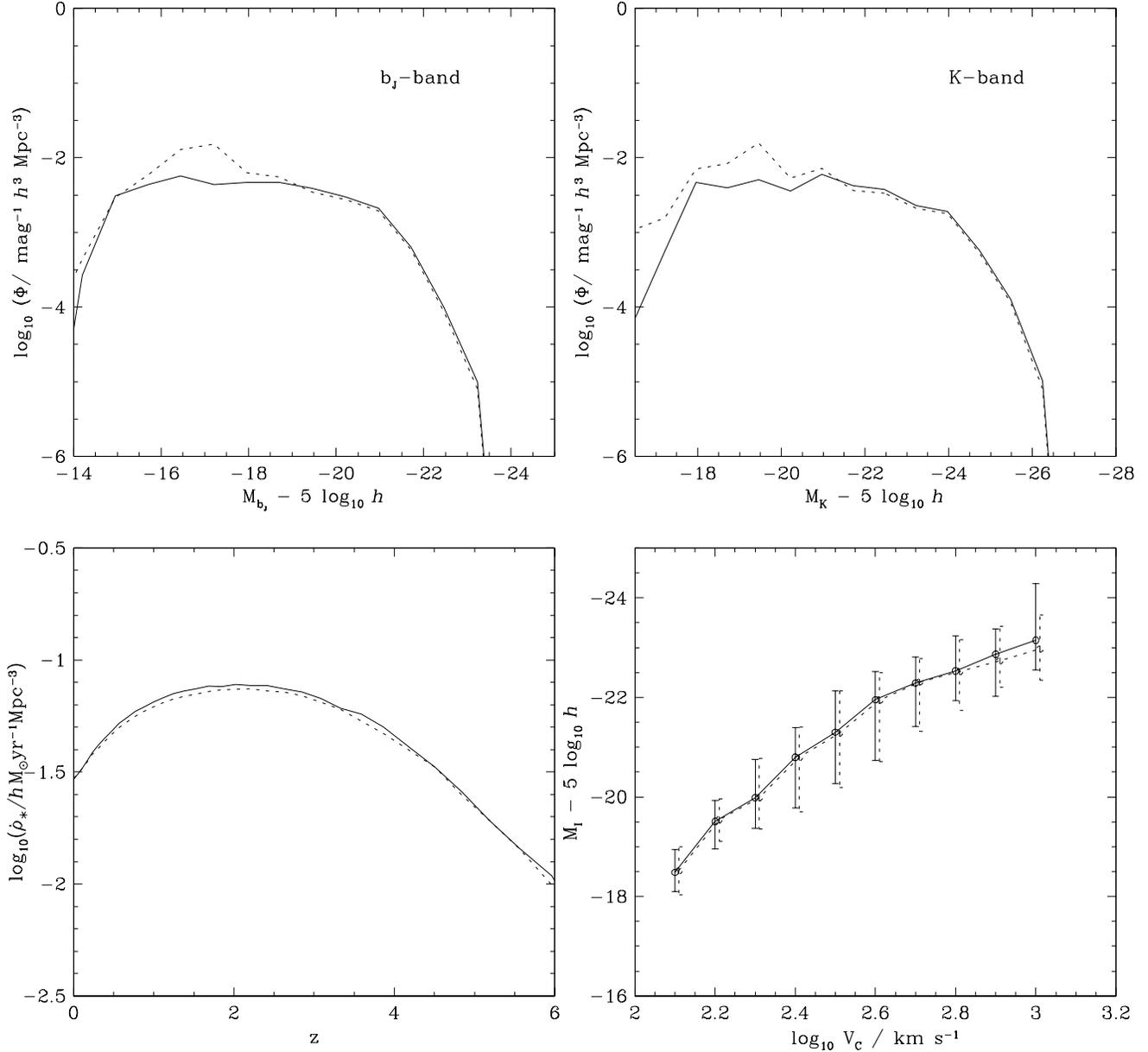}
\caption{Luminosity functions, star formation histories and Tully-Fisher relations for galaxies predicted by the N-body \galf\ model using merger trees obtained from the GIF simulation with two different methods of enforcing mass conservation. The solid lines show results obtained when mass conservation in the merger trees is enforced by increasing the masses of halos less massive than their progenitors. The dotted lines show the results obtained if, instead, the masses of the progenitors of such halos are reduced.}
\label{fig:mass_lftf}
\end{figure*}

The upper panels of Fig.~\ref{fig:mass_lftf} show the galaxy luminosity functions in the $\rm{b_J}$ and K bands predicted by the N-body \galf\ model with the parameters of Section~\ref{sec:galparams}, using the two different methods described in Section~\ref{sec:mass} to enforce mass conservation in the merger trees.
Over most of the luminosity range plotted, the two curves are essentially identical but there appear to be more galaxies at very faint $\rm{b_J}$ magnitudes when mass is removed from the merger trees. The majority of these galaxies formed in halos near the 10 particle ($\simeq 1.4 \times 10^{11} h^{-1}\rm{M_{\odot}}$) mass resolution limit imposed by the FOF group finder and their halos subsequently merged with other, larger dark matter halos. When mass conservation is enforced by removing mass from the merger trees (the dotted lines in Fig.~\ref{fig:mass_lftf}) it is possible to end up with some halos with mass less than the resolution limit which can harbour galaxies with $\rm{b_J}$ band magnitudes around -14 or fainter. If, instead, mass is added to halos less massive than their progenitors, then the merger trees contain no halos with masses below the FOF resolution threshold and hence fewer faint galaxies.
 
These sub-resolution halos often exist in the merger trees of larger halos and could affect the evolution of larger, brighter galaxies. However, the agreement of the luminosity functions suggests that any effect is insignificant. The global star formation history and Tully-Fisher relation shown in the lower panels of Fig.~\ref{fig:mass_lftf} are similarly unaffected.

Overall, the choice of mass conservation method appears to make very little difference to the quantities plotted in Fig.~\ref{fig:mass_lftf}, which suggests that the small amounts of mass being added to or removed from the merger trees do not significantly affect the properties of the resulting galaxies. The only region of the luminosity function which is affected is largely populated by galaxies which formed in halos with little or no resolved merger history, where the model cannot be expected to give reliable results. For the remainder of this paper we choose to enforce mass conservation by adding mass to the merger trees since this does not introduce halos with masses below the resolution limit.

\subsection{Comparison with standard GALFORM}
\label{sec:standard-galform}

The mass resolution of the merger trees taken from the GIF simulation is equal to 10 particle masses or $1.4\times 10^{11} h^{-1}\rm{M_{\odot}}$, i.e. $N_{\rm{min}}=10$. This is much larger than the mass resolution, $M_{\rm{res}}=5.0 \times 10^{9} h^{-1}\rm{M_{\odot}}$, used by Cole \etal (\shortcite{cole2k}). This will clearly affect the properties of the galaxies predicted by the N-body \galf\ model, since gas will be unable to cool and start forming stars until lower redshifts when halos with masses greater than $M_{\rm{res}}$ have formed. In order to investigate the effect of limited mass resolution on the N-body \galf\ model, we identify the properties of the merger trees which differ between standard and N-body \galf\ and use this knowledge to produce a modified version of the standard \galf\ model which reproduces the behaviour of the N-body \galf\ model. We can then increase the mass resolution of the merger trees in the modified model and observe the effects on the predicted galaxy properties.

There are four main reasons why the merger trees in the two models may differ. Firstly, there is the difference in mass resolution described above. Therefore, we initially degrade the mass resolution of the standard \galf\ model to match that of the GIF simulation by setting the minimum halo mass, $M_{\rm{res}}$, equal to the mass of $(N_{\rm{min}}-1)$ dark matter particles --- any halo of this mass or less in the N-body simulation would not be identified by the FOF group finder and would not be included in the N-body merger trees.

Secondly, Jenkins \etal (\shortcite{j2001}) have shown that the Press \& Schechter (\shortcite{ps74}) halo mass function (used in the standard \galf\ model) differs somewhat from the mass function determined from N-body simulations. We replace the Press-Schechter mass function in the standard \galf\ model with the mass function determined by Jenkins \etal This ensures that the distribution of halo masses at $z=0$ in the standard \galf\ model matches the distribution in the simulation. 

The number of timesteps also differs between the two models. In the standard \galf\ model we use 150 timesteps evenly spaced in $\log_{10} (1+z)$, whereas in the N-body case we have only 44 simulation outputs. However, we find that if we degrade the time resolution of the standard \galf\ model to match that of the N-body model the properties of the galaxy populations predicted change very little.

Finally, the distribution of progenitor masses for halos of a given mass predicted by the standard \galf\ model does not reproduce the distribution found in N-body simulations with complete accuracy. Benson \etal (\shortcite{b2001a}) show that an empirical correction can be used to bring the progenitor mass distributions in the semi-analytic and N-body merger trees into closer agreement. The threshold linear overdensity for collapse from the spherical collapse model, $\delta_{c}$, is replaced with an effective threshold $\delta^{\rm{eff}}_c=f_{\delta_c}\delta_c$. In the $\Lambda\rm{CDM}$ cosmology employed in the GIF simulation, the following form for $f_{\delta_c}$ was found by Benson \etal to give reasonable agreement between the progenitor mass functions between redshifts 0 and 3:
\begin{equation} f_{\delta_c}=1+0.14[\log_{10}(\rm{M}_{\rm{halo}}/\it{h}^{-1}\rm{M_{\odot}})-15.64], \label{eqn:ajbfix} \end{equation} where $\rm{M}_{\rm{halo}}$ is the mass of the final halo at redshift $z=0$. This form of modification was suggested by Tormen (\shortcite{tormen98}). 

\begin{figure*}
\epsfxsize=18.0 truecm \epsfbox{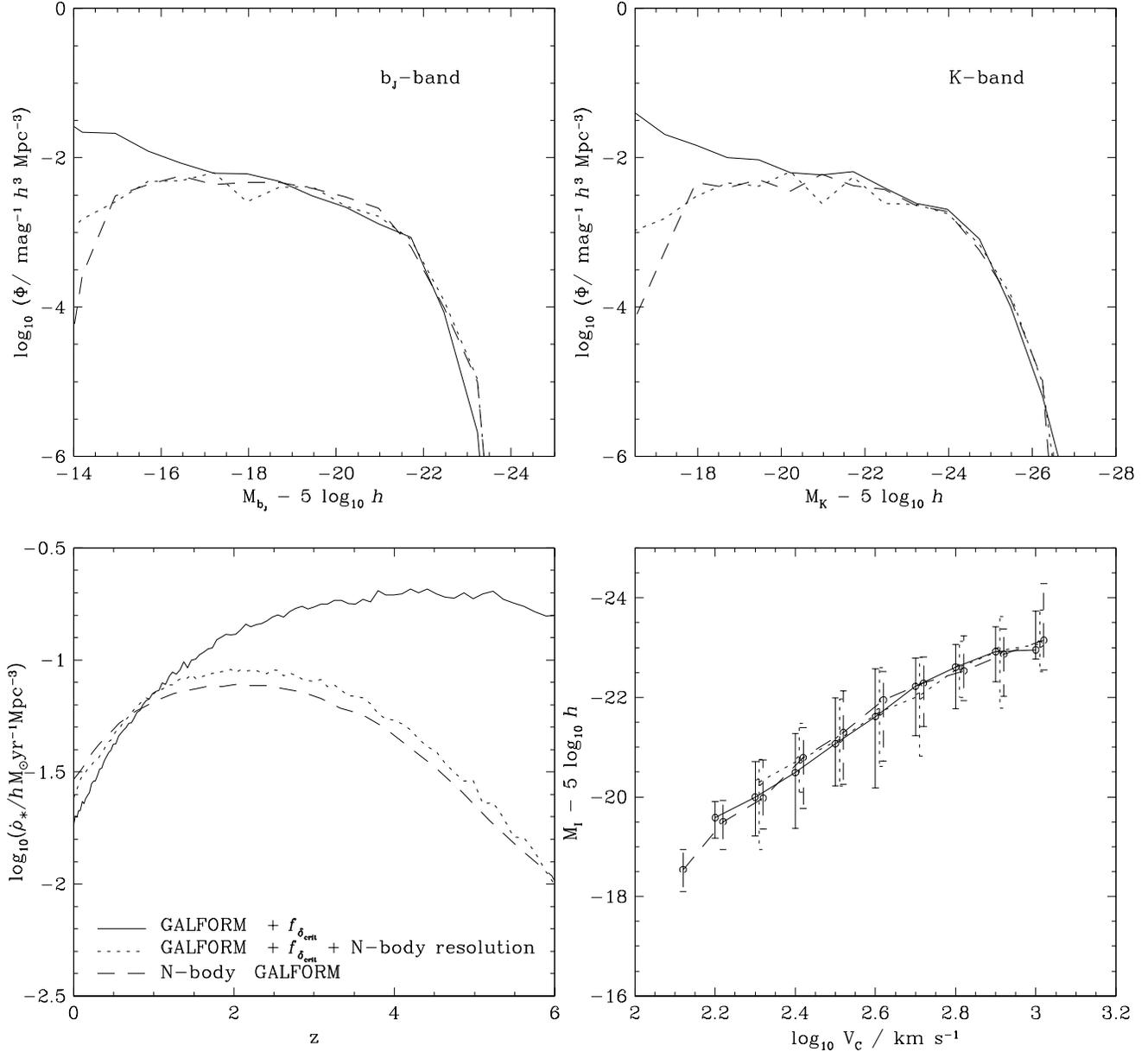}
\caption{Luminosity functions, star formation histories and Tully-Fisher relations for three different models. The solid lines correspond to the \galf\ model using Monte Carlo generated merger trees as described by Cole \etal (\shortcite {cole2k}), with the modifications explained in Section \ref{sec:nbody-galform} and a mass resolution of $5 \times 10^{9} h^{-1}\rm{M_{\odot}}$. The dotted lines show results from the same model with a mass resolution of $1.4 \times 10^{11} h^{-1}\rm{M_{\odot}}$, equivalent to that of the GIF simulation. The dashed lines show results obtained from the N-body \galf\ model which uses merger trees derived from the simulation.}
\label{fig:compare_res}
\end{figure*}

These modifications are intended to produce semi-analytic merger trees with statistical properties closely matched to those of the N-body merger trees. Fig.~\ref{fig:compare_res} shows the galaxy luminosity functions in the $\rm{b_J}$ and K bands, Tully-Fisher relations and global star formation histories for both the modified \galf\ model described above (dotted lines) and the N-body \galf\ model (dashed lines). It can be seen from the figure that these two models predict populations of galaxies with very similar statistical properties. The luminosity functions are in reasonable agreement for $\rm{K}$ brighter than about -18 and $\rm{b}_{\rm{J}}$ brighter than about -15. The Tully-Fisher relations and star formation histories are also in close agreement. 

As pointed out previously, the fainter galaxies in these models occupy halos with very poorly resolved merger histories and their properties may be largely determined by the effects of limited mass resolution. The solid lines in Fig.~\ref{fig:compare_res} show the properties of the galaxies in the modified \galf\ model when the minimum halo mass $M_{\rm{res}}$ is reduced to $5.0 \times 10^{9} h^{-1}\rm{M_{\odot}}$. This is much less massive than the smallest halo Benson \etal were able to resolve in their simulations and consequently, in this regime, eqn.~(\ref{eqn:ajbfix}) has not been tested and cannot be relied upon to produce a realistic distribution of progenitor masses. We do not expect this model to reproduce the results of Cole \etal but we show it only to provide some indication of the magnitude of the effect of introducing low mass halos into the merger trees.

This ``improvement'' in mass resolution increases the number of faint galaxies, which form in small, previously unresolved halos. With a higher minimum halo mass the gas in these small halos is unable to cool until it becomes incorporated into objects more massive than $M_{\rm{res}}$. This is reflected in the luminosity functions which show that there are slightly more bright galaxies and far fewer faint galaxies at $z=0$ in the model with poor mass resolution. The star formation history is consistent with this, showing that poor mass resolution results in reduced star formation at $z > 1$ and increased star formation at $z \approx 0$. However, calculating the global star formation rate involves a sum over all halos. At high redshifts this includes a large number of halos of low mass whose abundances may be unrealistic due to our extrapolation of eqn.~(\ref{eqn:ajbfix}). Reducing $M_{\rm{res}}$ appears to have little or no effect on the Tully-Fisher plot.

Overall, the predictions of the N-body \galf\ model closely match those of the standard \galf\ model when we take into account the differences in the halo mass function, the progenitor mass distribution and the mass resolution. The differences between the modified \galf\ models with high and low mass resolution indicate that, at low luminosities, the properties of the galaxies in the N-body model are seriously affected by the resolution of the simulation. In order to attempt accurately to reproduce the properties of observed galaxy populations with $\rm{b_J}$ band magnitudes fainter than about -17, an N-body simulation with significantly improved mass resolution would be required.

\section{Conclusions}

In this paper we have examined how the statistical properties of the galaxies predicted by a semi-analytic model depend on the way in which the dark matter halo merger histories are created. We have developed a method for calculating merger histories from N-body simulations and used the resulting merger trees in a semi-analytic model of galaxy formation based on that of Cole \etal (\shortcite{cole2k}). We refer to this model as N-body \galf\, and compare it to an otherwise identical ``standard \galf\ '' model, which uses halo merger histories generated using the Monte-Carlo algorithm of Cole \etal This algorithm is based on the extended Press-Schechter (EPS) theory.

We find that in a significant number of cases, halos in the N-body merger trees are less massive than their progenitors at the previous timestep. When this happens we are forced artificially to adjust the masses of the halo or its progenitors, since in our semi-analytic galaxy formation model halos may not lose mass.  However, the luminosity function, Tully-Fisher relation and global star formation history of the galaxies predicted by the semi-analytic model remain almost exactly the same whether we add mass to the halo or remove mass from the progenitors when we encounter this problem. We conclude that the changes we are forced to make to the halo masses have very little effect on the semi-analytic model.

If the mass resolution in the standard \galf\ model is degraded to that of the N-body simulation and the empirical fit of Benson \etal (\shortcite{b2001a}) is used to correct the distribution of halo progenitor masses, we obtain luminosity functions and Tully-Fisher relations in very good agreement with the N-body \galf\ model. This shows that, apart from the issue of mass resolution, the only significant statistical differences between the N-body merger trees and those of Cole \etal are due to the known discrepancy between EPS theory and the results of N-body simulations.
 
 By improving the mass resolution in the standard \galf\ model to that used by Cole \etal we were able to obtain an indication of the effects of limited mass resolution on the N-body model. The mass resolution in the N-body merger trees is imposed by the particle mass in the GIF simulation, since halos with fewer than 10 particles ($1.4 \times 10^{11} \it{h}^{-1} \rm{M}_{\odot}$) are not resolved. This limitation has a noticeable effect on the galaxy luminosity function and we find slightly more very bright galaxies, since gas may only cool in resolved halos. If only massive halos are resolved, cooling is delayed resulting in brighter galaxies at $z=0$. However, the most obvious effect of poor mass resolution is a drastic reduction in the number of galaxies with $b_{\rm{J}}$ magnitudes fainter than about -17. This demonstrates that the mass resolution of the GIF simulation is insufficient to make reliable predictions at these magnitudes. At brighter magnitudes the luminosity functions remain in good agreement. 

In conclusion, when used as the starting point for semi-analytic modelling of galaxy formation, merger trees taken from an N-body simulation using the technique described in this paper result in similar galaxy populations to those obtained using the (slightly modified) Monte-Carlo algorithm of Cole \etal This supports the reliability of our method and provides a means to populate large cosmological N-body simulations with semi-analytic galaxies at a fraction of the computational cost of a hydrodynamic simulation of the same volume. When applied to the dark matter component of an Smooth Particle Hydrodynamics (SPH) simulation, our model will also allow us to compare SPH and semi-analytic treatments of galaxy formation, and in particular the cooling of gas within halos, on a halo-by-halo basis. This comparison is reported in a companion paper (Helly \etal \shortcite{paperII}.)


\label{sec:conclusions}
\section*{Acknowledgements}
We acknowledge support from PPARC and the Royal Society.


\begin{thebibliography}{}

\bibitem[\protect\citename{Benson \etal}2000]{b2000}
Benson, A. J., Cole, S., Frenk, C. S., Baugh, C. M., Lacey, C. G., 2000, MNRAS, 311, 793

\bibitem[\protect\citename{Benson \etal}2001a]{b2001a}
Benson, A.J., Pearce, F.R., Frenk, C.S., Baugh, C.M., Jenkins, A., 2001, MNRAS, 320, 261

\bibitem[\protect\citename{Benson \etal}2001b]{b2001b}
Benson, A.J., Frenk, C.S., Baugh, C.M., Cole, S., Lacey, C.G., 2001, MNRAS, 327, 1041

\bibitem[\protect\citename{Bond \etal}1991]{bond91}
Bond, J.R., Cole, S., Efstathiou, G., Kaiser, N., 1991, ApJ, 379, 440

\bibitem[\protect\citename{Bower }1991]{bower91}
Bower, R.G., 1991, MNRAS, 248, 332



\bibitem[\protect\citename{Cole }1991]{cole91}
Cole, S., 1991, ApJ, 367, 45

\bibitem[\protect\citename{Cole \etal}1994]{cole94}
Cole, S., Aragon-Salamanca, A., Frenk, C.S., Navarro, J.F., Zepf, S.E., 1994, MNRAS, 271, 781 

\bibitem[\protect\citename{Cole \& Lacey }1996]{cl96}
Cole, S., Lacey, C.G., 1996, MNRAS, 281, 716

\bibitem[\protect\citename{Cole \etal}2000]{cole2k} 
Cole S., Lacey C.G., Baugh C.M., Frenk C.S., 2000, MNRAS, 319, 168

\bibitem[\protect\citename{Couchman, Thomas \& Pearce }1995]{ctp95}
Couchman, H.M.P., Thomas, P.A., Pearce, F.R., 1995, ApJ, 452, 797

\bibitem[\protect\citename{Davis \etal}1985]{davis85} 
Davis, M., Efstathiou, G., Frenk C.S., White, S.D.M, 1985, ApJ, 292, 371

\bibitem[\protect\citename{Eke, Cole \& Frenk }1996]{ecf96}
Eke, V.R., Cole, S.M., Frenk, C.S., 1996, MNRAS, 282, 263




\bibitem[\protect\citename{Governato \etal}1998]{governato98}
Governato, F., Baugh, C. M., Frenk, C. S., Cole, S., Lacey, C. G., Quinn, T., Stadel, J., 1998, Nature, 392, 359

\bibitem[\protect\citename{Governato \etal}1999]{governato99}
Governato, F., Babul, A., Quinn, T., Tozzi, P., Baugh, C. M., Katz, N., Lake, G., 1999, MNRAS, 307, 949

\bibitem[\protect\citename{Gross \etal}1998]{gross98}
Gross, M.A.K., Somerville, R.S., Primack, J.R., Holtzman, J., Klypin, A., 1998, MNRAS, 301, 81

\bibitem[\protect\citename{Helly \etal}2002]{paperII}
Helly, J.C., Cole, S.M., Frenk, C.S., Baugh, C.M., Benson, A., Lacey, C.G., Pearce, F.R., 2002, submitted to MNRAS

\bibitem[\protect\citename{Jenkins \etal}2001]{j2001}
Jenkins, A., Frenk, C.S., White, S.D.M., Colberg, J.M., Cole, S., Evrard, A.E., Couchman, H.M.P., Yoshida, N., 2001, MNRAS, 321, 372 

\bibitem[\protect\citename{Jenkins \etal}1998]{j98}
Jenkins, A., Frenk, C. S., Pearce, F. R., Thomas, P. A., Colberg, J. M., White, S. D. M., Couchman, H. M. P., Peacock, J. A., Efstathiou, G., Nelson, A. H., 1998, ApJ, 499, 20

\bibitem[\protect\citename{van Kampen \etal}1999]{vk99}
van Kampen, E., Jimenez, J., Peacock, J.A., 1999, MNRAS, 310, 43 



\bibitem[\protect\citename{Kauffmann \etal}1999]{kauffmann99} 
Kauffmann, G., Colberg, J.M., Diaferio, A., White, S.D.M., 1999, MNRAS, 303, 188

\bibitem[\protect\citename{Kauffmann \& Haehnelt }2000]{kh2000} 
Kauffmann, G., Haehnelt, M., 2000, MNRAS, 311, 576


\bibitem[\protect\citename{Lacey }2002]{lacey02}
Lacey, C.G., Baugh, C.M., Frenk, C.S, Cole, S.M., Bressan, S., Granato, G.L., Silva, L., 2002, RevMexAA Serie de Conferencias, in preparation

\bibitem[\protect\citename{Lacey \& Cole }1993]{lc93} 
Lacey, C.G., Cole, S., 1993, MNRAS, 262, 627

\bibitem[\protect\citename{Lacey \& Cole }1994]{lc94} 
Lacey, C.G., Cole, S., 1994, MNRAS, 271, 676

\bibitem[\protect\citename{Lacey \& Silk }1991]{ls91} 
Lacey, C.G., Silk, J., 1991, ApJ, 381, 14


\bibitem[\protect\citename{Mo \& White }2002]{mw02}
Mo, H.J., White, S.D.M., 2002, MNRAS, 336, 112

\bibitem[\protect\citename{Navarro, Frenk \& White }1995]{nfw95}
Navarro, J.F., Frenk, C.S., White, S.D.M., 1995, MNRAS, 275, 720

\bibitem[\protect\citename{Navarro, Frenk \& White }1996]{nfw96}
Navarro, J.F., Frenk, C.S., White, S.D.M., 1996, ApJ, 462, 563

\bibitem[\protect\citename{Navarro, Frenk \& White }1997]{nfw97}
Navarro, J.F., Frenk, C.S., White, S.D.M., 1997, ApJ, 490, 493



\bibitem[\protect\citename{Pearce \& Couchman }1997]{pc97}
Pearce, F.R., Couchman, H.M.P., 1997, New Astronomy, 2 ,411



\bibitem[\protect\citename{Press \& Schechter }1974]{ps74}
Press, W.H., Schechter, P., 1974, ApJ, 187, 425


\bibitem[\protect\citename{Sheth \& Tormen }2001]{st01}
Sheth, R. K., Mo, H.J., 2001, MNRAS, 323, 1

\bibitem[\protect\citename{Somerville \& Primack }1999]{sp99}
Somerville, R.S., Primack, J.R., 1999, MNRAS, 310, 1087




\bibitem[\protect\citename{Sutherland \& Dopita }1993]{sd93}
Sutherland, R., Dopita, M., 1993, ApJS, 88, 253

\bibitem[\protect\citename{Tormen }1997]{tormen97}
Tormen, G., 1997, MNRAS, 300, 773

\bibitem[\protect\citename{Tormen }1998]{tormen98}
Tormen, G., 1998, MNRAS, 297, 648

\bibitem[\protect\citename{Wechsler \etal}2001]{wechsler2001}
Wechsler, R. H., Somerville, R. S., Bullock, J. S., Kolatt, T. S., Primack, J. R., Blumenthal, G. R., Dekel, A., 2001, ApJ, 554, 85


\bibitem[\protect\citename{White \& Frenk }1991]{wf91}
White, S.D.M., Frenk, C.S., 1991, ApJ, 379, 52


\end{thebibliography}
\end{document}